\documentclass[preprint,showpacs,preprintnumbers,amsmath,amssymb]{revtex4}

\usepackage{graphicx}
\usepackage{dcolumn}
\usepackage{bm}

\begin{document}


\title{Direct laser cooling Al$^+$ ions optical clocks}

\author{J. Zhang}
\author{K. Deng}
\author{J. Luo}
\author{Z. H. Lu}
\email{zehuanglu@mail.hust.edu.cn}
\affiliation{
MOE Key Laboratory of Fundamental Physical Quantities Measurement,
School of Physics, Huazhong University of Science and Technology,
Wuhan, 430074, China
}

\date{\today}

\begin{abstract}
Al$^+$ ions optical clock is a very promising optical frequency standard candidate due to its extremely small blackbody radiation shift. It has been successfully demonstrated with indirect cooled, quantum-logic-based spectroscopy technique. Its accuracy is limited by second-order Doppler shift, and its stability is limited by the number of ions that can be probed in quantum logic processing. We propose a direct laser cooling scheme of Al$^+$ ions optical clocks where both the stability and accuracy of the clocks are greatly improved. In the proposed scheme, two Al$^+$ ions traps are utilized. The first trap is used to trap a large number of Al$^+$ ions to improve the stability of the clock laser, while the second trap is used to trap a single Al$^+$ ions to provide the ultimate accuracy. Both traps are cooled with a continuous wave 167 nm laser. The expected clock laser stability can reach $9.0\times10^{-17}/\sqrt{\tau}$. For the second trap, in addition to 167 nm laser Doppler cooling, a second stage pulsed 234 nm two-photon cooling laser is utilized to further improve the accuracy of the clock laser. The total systematic uncertainty can be reduced to about $1\times10^{-18}$. The proposed Al$^+$ ions optical clock has the potential to become the most accurate and stable optical clock.
\end{abstract}

\pacs{06.30.Ft, 37.10.Ty, 32.70.Jz}

\maketitle

Atomic clocks have many applications in precision spectroscopy \cite{Hall,Hansch}, test of relativity \cite{Schiller,Wolf,Chou}, and astronomy \cite{Steinmetz,Li}. They also have significant impact on applications found in everyday life, such as global navigation systems, and high-speed data transmission and communication. Given the increasing demand on better frequency standards for diverse applications, for example in measuring the time variations of fundamental constants \cite{Rosenband, Godun, Huntemann2}, more stable and accurate atomic clocks are needed. Optical clocks, or atomic clocks using optical transitions, have much higher $Q$ value in comparison to that of microwave atomic clocks. As a consequence, optical clocks potentially can have better stability and accuracy than today's best microwave atomic clocks. Currently the best cesium fountain clocks have fractional frequency uncertainties of $1.1\times10^{-16}$ \cite{Heavner}, while optical clocks have long been recognized to have the potential to reach a systematic fractional frequency uncertainty below $10^{-18}$.

Many groups are working on optical clocks with the goal of realizing improved frequency standards for a new definition of the SI second and testing fundamental physics. Currently there are two possible routes to develop an optical clock. One is based on trapped ions in a Paul trap, the other is based on neutral atoms trapped in an optical lattice. Potential choices of single ion species include Hg$^+$ \cite{Oskay}, Sr$^+$ \cite{Barwood, Dube}, Yb$^+$ \cite{King, Huntemann}, Ca$^+$ \cite{Hashimoto, Huang}, In$^+$ \cite{Wang}, and Ba$^+$ \cite{Fortson}, while potential choices of optical lattice clocks include Sr \cite{Katori, Paris, Bloom}, Yb \cite{Ludlow}, and Hg \cite{Bize}.

The 3s$^2$ $^1$S$_0$ - 3s3p $^3$P$_0$ transition in Al$^+$ at 267.4 nm, which has a narrow natural line width of 8 mHz, is an excellent choice for optical clock transition. The $Q$ value of this transition is $1.4\times10^{17}$. Its electric quadrupole shift is negligible since the transition is between J = 0 states, therefore the transition frequency is not influenced by the electric field gradient at the center of the ion trap. Furthermore, blackbody radiation contribution to the fractional frequency uncertainty of the Al$^+$ ion clock is as low as $4\times10^{-19}$ at room temperature, which is the smallest among all atomic species currently under consideration for optical clocks \cite{Safronova}. As a consequence, Al$^+$ ion clocks have very small systematic uncertainty.

The stability of an optical clock is limited by quantum project noise (QPN). QPN sets the quantum limit to a clock stability, and it can be expressed as \cite{Ye}
\begin{equation}
\sigma(\tau)=\frac{\eta}{\pi Q}\sqrt{\frac{T_c}{N\tau}}.
\end{equation}
Here $Q$ is the quality factor, $T_c$ is the clock cycle time, $N$ is the number of atoms, $\tau$ is the averaging time, and $\eta$ is a numerical factor near unity. In the case of Al$^+$ ion clock, $N=1$. In a typical neutral atoms optical lattice clock, $N$ is on the order of $10^3$. So it is generally accepted that optical lattice clocks have better stability, while single ion optical clocks have better accuracy, although this belief has been shaken recently with the latest neutral atoms optical lattice clock results \cite{Bloom, Ludlow}.

In this Letter, we suggest an Al$^+$ ion optical clock scheme that has the potential of surpassing both the stability and accuracy performance levels of current best optical clocks. In this scheme, two Al$^+$ ion traps are utilized. One trap is used to trap large number of Al$^+$ ions (at the order of $10^4$), the other trap is used to trap single Al$^+$ ions. Figure 1(a) shows the working principle of the suggested scheme. The frequency of the 267.4 nm laser (the clock laser) is first locked to the atomic transition of Al$^+$ ions in the first trap. Due to a large number of Al$^+$ ions involved, the stability of the locked clock laser can be greatly improved to even less than that of an optical lattice clock. The clock laser is then further locked to the second ion trap that traps only one single Al$^+$ ion. In this way, the accuracy of the clock laser is limited by the single ion trap accuracy. This optical clock approach might provide an ultimate optical clock with a stability surpassing that of an optical lattice clock and an accuracy of a single ion optical clock.

In order to explain the potential improvement of the new proposal, it is helpful to review the performance limitation of current Al$^+$ ion optical clock. The idea of Al$^+$ ion optical clock is first proposed by Yu, Dehmelt, and Nagourney \cite{Yu}. In order to construct an Al$^+$ ion optical clock, at least a cooling laser is needed to cool the ion in the trap and detect its quantum state through the method of electron-shelving. The fluorescence rate of the cooling transition shall be large enough to facilitate efficient cooling and detection. In Al$^+$ ion the transition which meets the requirement is from 3s$^2$ $^1$S$_0$ to 3s3p $^1$P$_1$. Its natural linewidth is 230 MHz and its wavelength is 167 nm. However, since there is no continuous wave (CW) laser which can reaches this wavelength at the time of the proposal, the Al$^+$ clock cannot be experimentally realized after it was proposed. In order to overcome this difficulty, a method of quantum logic spectroscopy (QLS) was proposed to cool the ion and detect the clock transition \cite{Schmidt}. The QLS method can be implemented in a linear trap, where an auxiliary (for example, Be$^+$) logic ion together with an Al$^+$ ion are trapped \cite{Wineland1}. The logic ion provides sympathetic laser cooling, state initialization, and state detection for the simultaneously trapped Al$^+$ ion. After sympathetic laser cooling, both the logic ion and the Al$^+$ ion can be cooled to vibrational ground state. The Coulomb interaction of the ions couples their motions. Then the state detection is achieved through a coherent transfer of the Al$^+$ ion's internal quantum state onto the logic ion. Through the QLS method, the first Al$^+$ optical clock was successfully demonstrated at the National Institute of Standards and Technology (NIST) \cite{Wineland1}. Its frequency was compared to a Hg$^+$ ion optical clock, the comparison results yield a preliminary constraint on the temporal variation of the fine-structure constant of $(-1.6\pm2.3)\times10^{-17}$/year \cite{Rosenband}. In subsequent experiments, Mg$^+$ ion was used as the logic ion to realize an Al$^+$ ion clock with better systematic uncertainty \cite{Chou2}. Using two Al$^+$ ion clocks, NIST verified the relativistic time dilation effect and the gravitational time dilation effect \cite{Chou}.

The systematic uncertainty of the QLS-based Al$^+$ ion clock ($8\times10^{-18}$) is dominated by the second order Doppler effects, which is mainly due to the fact that two ions have to be trapped simultaneously. If only a single aluminum ion is trapped in the trap, it can be expected that the systematic uncertainty can be greatly reduced. In order to trap a single Al$^+$ ion, as mentioned above, a CW 167 nm laser has to be built for direct laser cooling. One of the possible schemes to build a 167 nm laser is to use two frequency doubling processes, starting from a 667 nm laser \cite{Chi}. Frequency doubling from 667 nm to 334 nm can be realized by a second harmonic generation (SHG) cavity with a BBO crystal. However, frequency doubling from 334 nm to 167 nm cannot be realized without a suitable nonlinear crystal. This problem can be solved after the invention of Potassium Fluoroboratoberyllate (KBBF) crystal \cite{KBBF1}. With KBBF's absorption edge at 153 nm, the lowest obtained wavelength is 156 nm under pulsed operation \cite{KBBF2}. In principle, KBBF crystal can be used to generate a 167 nm CW laser through direct SHG as long as enough length of crystal can be grown. The growth technique of KBBF crystal has made significant progress, overcoming the early growth problems. As a consequence, KBBF crystal with length more than 6 mm can be grown \cite{KBBF1}. Recently, a $1.3$ mW CW 191 nm laser using SHG of KBBF crystal has been demonstrated \cite{KBBF3}, thus laying a solid foundation for a CW 167 nm laser with tens of microwatts power output. Since the saturation power of Al$^+$ ion $^1$S$_0$ to $^1$P$_1$ transition is only 5 $\mu$W assuming a focused spot radius of 5 $\mu$m, this power level shall be enough to fulfill the cooling laser requirements.

In this proposal, we will evaluate the systematic uncertainty of an Al$^+$ ion optical clock assuming a 167 nm laser is available. In this case, we only need to trap one single Al$^+$ ion, and we first Doppler cool the ion with the 167 nm laser. Figure 1(b) shows the relevant energy levels of Al$^+$ ions. Two types of residual motions cause time-dilation shifts: excess micromotion and secular motion. In both cases the relative clock frequency shifts by
\begin{equation}
\frac{\Delta\nu}{\nu_0}=-\frac{<v^2>}{2c^2},
\end{equation}
where $<v^2>$ is the mean-squared velocity of Al$^+$ ions. Here we have neglected the frequency-dependent term that corresponds to the Stark shift from the motion-inducing electric fields. Stark shift correction is less than 2\% for a motional frequency of tens of MHz \cite{Berkeland}.

After the 167 nm laser Doppler cooling is finished, the second order secular motion Doppler shift can be calculated as
\begin{equation}
\frac{\Delta\nu}{\nu_0}=-\frac{3\hbar\gamma}{4mc^2}.
\end{equation}
Here $m$ is the mass of Al$^+$ ions, $\gamma$ is the natural line width of the transition from $^1$S$_0$ to $^1$P$_1$. The calculated fractional frequency shift is $-2.8\times10^{-17}$. The second order micromotion Doppler shift is given by \cite{Berkeland}
\begin{equation}
\frac{\Delta\nu}{\nu_0}=-\frac{1}{4}(\frac{\beta \Omega}{2\pi\nu_0})^2.
\end{equation}
$\beta$ is a parameter characterizing the magnitude of excess micromotion. For a conservative trap parameter of $\beta = 0.1$ \cite{Keller}, and an ion trap RF driving frequency $\Omega$ of 20 MHz, the fractional frequency shift is $-3.0\times10^{-19}$.

From the above calculation we can see that just using 167 nm direct laser cooling, the second-order Doppler shift due to secular motion is higher than that of QLS-based method ($-1.6\times10^{-17}$) \cite{Chou2}. A straightforward way to solve this problem is to introduce a second stage cooling based on the spin-forbidden intercombination transition from $^1$S$_0$ to $^3$P$_1$ through a 267 nm laser \cite{Katori2, Ertmer}. Unfortunately, the $^1$S$_0$ to $^3$P$_1$ transition natural line width is only 500 Hz, and there is no suitable quenching transition for Al$^+$ ion \cite{Ertmer2, Garreau, Curtis, Sterr}. As a consequence, the 267 nm laser cooling rate is too small to be effective.

To solve this problem, we propose a second stage laser cooling based on two-photon transition. Two-photon cooling with CW lasers has been proposed and demonstrated in neutral atomic systems before \cite{Cruz, Thomsen}. In our case, the relevant two-photon transition is from $^1$D$_2$ (state $\left|2\right\rangle$) to $^1$S$_0$ (state $\left|0\right\rangle$) with a transition rate $\Gamma_{20}$ of $8.37\times10^3$ s$^{-1}$. The corresponding two-photon transition laser wavelength is 234 nm. This value does not seem to be too large either, but there are another two-step transition process from $\left|2\right\rangle$ decays to $\left|0\right\rangle$ through $^1$P$_1$ (state $\left|1\right\rangle$). The relevant transition rates are $\Gamma_{21}=4.8\times10^5$ s$^{-1}$ and $\Gamma_{10}=1.41\times10^9$ s$^{-1}$, respectively. The combined transition rate from $\left|2\right\rangle$ to $\left|0\right\rangle$ is $\Gamma_2=\Gamma_{20}+\Gamma_{eff}$, where $\Gamma_{eff}$ can be expressed as \cite{Thomsen}
\begin{equation}
\Gamma_{eff}=\Gamma_{21}+(\Gamma_{21}+\Gamma_{10})\frac{\Omega_{10}^2/2}{(\Gamma_{21}+\Gamma_{10})^2/2+\Delta_{10}^2}.
\end{equation}
Here $\Omega_{10}$ and $\Delta_{10}$ are Rabi frequency and detuning to state $\left|1\right\rangle$, respectively. The transition rate from $\left|2\right\rangle$ to $\left|0\right\rangle$ is dominated by $\Gamma_{21}$ so that $\Gamma_2\approx\Gamma_{21}=4.8\times10^5$ s$^{-1}$.

It can be calculated that the required saturation power for a CW 234 nm laser is over 200 mW, assuming a focused spot radius of 5 $\mu$m. Such power level is difficult to reach at the moment. Unless an enhancement cavity is used inside the ion trap vacuum chamber with associated other experimental issues. A simpler method is to use an ultrafast pulsed laser with center wavelength at 234 nm \cite{Kielpinski}. It can be calculated that the two-photon cooling saturation power for a 250 MHz repetition rate ultrafast laser is less than 100 mW, which is well within reach of current technological level.

With two-photon pulsed laser cooling as the second stage cooling, we can cool Al$^+$ ions to vibrational ground state in the strong confinement regime. In this condition, the second order Doppler shift is recoil limited. Second-order secular motion Doppler shift due to 234 nm pulsed laser is given by
\begin{equation}
\frac{\Delta\nu}{\nu_0}=-\frac{3h^2\nu_0^2}{2m^2c^4}.
\end{equation}
The result is $-6.5\times10^{-20}$. Second-order excess micromotion Doppler shift due to 234 nm pulsed laser is $-1.5\times10^{-19}$ according to Eq. (4).

In addition, motional heating of trapped ions can increase the second-order Doppler shift when ions are not cooled. The rate of increasing can be written as
\begin{equation}
\frac{\partial}{\partial t}\left(\frac{\Delta\nu}{\nu_0}\right)=-\frac{\dot{\bar{n}} \hbar\omega_s}{mc^2},
\end{equation}
where $\dot{\bar{n}}$ is the heating rate, and $\omega_s$ is the secular motion frequency. Taking $\dot{\bar{n}}=1$ s$^{-1}$, $\omega_s/2\pi=20$ MHz, and probe time of 1 s, the second-order Doppler shift due to this motional heating is $-1.6\times10^{-19}$.

The systematic shifts and uncertainties of the direct cooled Al$^+$ ion optical clocks are listed in table \ref{Table}. We use similar experimental conditions as those used in the QLS-based Al$^+$ ion optical clocks, with potentials for further improvements. It can be seen that the systematic uncertainties are reduced by a factor of eight in comparison to that of the QLS-based Al$^+$ ion optical clocks, and has the potential to become one of the most accurate optical clocks.

The stability of QLS-based Al$^+$ ion optical clocks is reported to be $2.0\times10^{-15}/\sqrt{\tau}$ assuming equal contributions of two comparison clocks \cite{Wineland1}. To increase the stability of the Al$^+$ ion optical clocks, two Al$^+$ ion traps are utilized. The first trap is used to trap a large number of Al$^+$ ions, and the second trap is the one that we just discussed to trap single Al$^+$ ions. The frequency of the 267.4 nm clock laser is first locked to the atomic transition of Al$^+$ ions in the first trap for faster reduction of the phase noise of the clock laser. Since we are not concerning about the accuracy for the first trap, a first stage direct 167 nm laser cooling is enough.

Similar ideas have been proposed before, where the frequency of the clock laser is locked to the atomic transition using several ensembles of atoms \cite{Sorensen, Rosenband2}. The stability of the clock laser scales as $\sqrt{\gamma}(\gamma T_1N)^{-m/2}$. Here $m$ is the number of ensembles, and each ensemble has equal number of atoms $N$. $\gamma$ is a parameter characterizing the frequency fluctuations of the free running clock laser, and $T_1$ is the probing time of the first trap.

In our scheme, we use two traps with unequal number of ions. Following the discussion of Ref.\cite{Sorensen}, the first ion trap is operated with a cycle time of $T_1$, and the second trap is operated with a cycle time of $T_2$, where $T_2$ is an integer multiples of $T_1$. The stability of the clock laser at time $T_2$ after locking to the first trap is
\begin{equation}
\sigma_1(T_2)=\frac{1}{\omega}\sqrt{\frac{1}{NT_1T_2}},
\end{equation}
where $\omega$ is the angular frequency of the clock laser. This stability is further improved after locking to the second trap,
\begin{equation}
\sigma_2(\tau)=\frac{1}{\omega}\sqrt{\frac{1}{\tau T_2}},
\end{equation}
where $\tau > T_2$.

The longest $T_2$ we can allow is determined by how well the clock laser is stabilized by the first trap as $T_{2,max}=\beta_2 NT_1$. Similarly, the longest $T_1$ is limited by the free running clock laser stability as $T_{1,max}=\beta_1/\gamma$. $\beta_1$ and $\beta_2$ are in the order of 0.1 \cite{Sorensen}. Then the clock laser stability can be expressed as
\begin{equation}
\sigma(\tau)=\frac{1}{\omega}\sqrt{\frac{\gamma}{\tau N\beta_1\beta_2}}.
\end{equation}
This result shall be compared with the clock laser stability when it is locked to a single ion trap,
\begin{equation}
\sigma(\tau)=\frac{1}{\omega}\sqrt{\frac{1}{\tau T_1}}=\frac{1}{\omega}\sqrt{\frac{\gamma}{\tau \beta_1}}.
\end{equation}

As an example, suppose we can trap $10^4$ ions in the first trap, and we take $\beta_1=\beta_2=0.1$, then our scheme can reach a stability level of $9.0\times10^{-17}/\sqrt{\tau}$, a factor of more than 20 times improvement in comparison with that of single ions trap. The expected stability improvement is shown in Fig. 2. This stability level will even surpass the stability level of optical lattice clocks. At this stability level, the free running clock laser Dick effect has to be taken into account. In Fig. 2, the Dick effect contribution of a currently best 40 mHz line width laser is also shown \cite{Kessler}.

There are questions that with an increased excess micromotion associated with $10^4$ ions in the first trap, whether a stability of $9.0\times10^{-17}/\sqrt{\tau}$ is achievable. For a linear quadrupole trap, the line broadening effect due to the excess micromotion can be calculated as \cite{Prestage1}
\begin{equation}
\Delta \nu=\frac{e^2}{8\pi\epsilon_0 m c^2}\frac{N}{L}\nu_0=32.1\ \text{Hz},
\end{equation}
where we have taken the trap length $L$ to be 1 cm. According to Eq. (1), the expected stability limit is $8.9\times10^{-17}/\sqrt{\tau}$, which is smaller than $9.0\times10^{-17}/\sqrt{\tau}$. Even better stability limit can be expected if we choose to use a multi-pole linear ion trap, for instance, at least a factor of ten improvement has been achieved with a 12-pole linear ion trap \cite{Prestage2}.

In conclusion, we propose an Al$^+$ ion optical clock scheme where two ion traps are utilized. The first trap is used to trap a large number of Al$^+$ ions to improve the stability of the clock laser, while the second trap is used to trap a single Al$^+$ ion to provide the ultimate accuracy. Both traps are cooled with a CW 167 nm laser. With the first trap trapping over $10^4$ ions, the expected clock laser stability can reach $9.0\times10^{-17}/\sqrt{\tau}$. Since trapping of $10^4$ ions is not a difficult task, with potential of trapping $10^6$ ions \cite{Drewsen}, the calculated stability can be treated as a conservative estimation. For the second trap, in addition to CW 167 nm laser Doppler cooling, a second stage pulsed 234 nm laser two-photon cooling is utilized to further improve the accuracy of the clock laser. The second order Doppler frequency shift can be reduced to a level of $10^{-19}$, and the total systematic uncertainty can be reduced to about $1\times10^{-18}$, which is greatly improved over the current QLS-based Al$^+$ ion optical clocks. We shall further emphasize that the proposal principle is not restricted to Al$^+$ ions clocks, but is applicable to other ions clocks.

\begin{acknowledgments}
We wish to thank H. Che for preliminary calculation of the two-photon transition rate. The project is partially supported by the National Basic Research Program of China (Grant No. 2012CB821300), the National Natural Science Foundation of China (Grant No. 91336213, 11174095, 11304109, and 91536116), and Program for New Century Excellent Talents by the Ministry of Education (NCET-11-0176).
\end{acknowledgments}

\newpage
\begin{table}[htbp]
\caption{Clock uncertainty budget.}
\label{Table}
\renewcommand\arraystretch{1.5}
\begin{center}
\begin{tabular}{cccccc}\hline

 &Fractional shift ($10^{-18}$)   &Uncertainty ($10^{-18}$)     \\\hline
Excess micromotion              &-0.15    &0.05     \\
Secular motion               &-0.07    &0.02     \\
Motional heating             &-0.16    &0.05    \\
Blackbody radiation shift              &1.2    &0.4     \\
Quadratic Zeeman shift               &-1079.9    &0.7     \\
Linear Doppler shift               &0    &0.3     \\
Clock laser Stark shift               &0    &0.2     \\
Background-gas collisions               &0    &0.5     \\
AOM frequency error               &0    &0.2     \\\hline
Total               &-1079.3    &1.03     \\\hline

\end{tabular}
\end{center}
\end{table}

\newpage

\begin{figure}[tfb]
\centerline{\includegraphics[width=14cm]{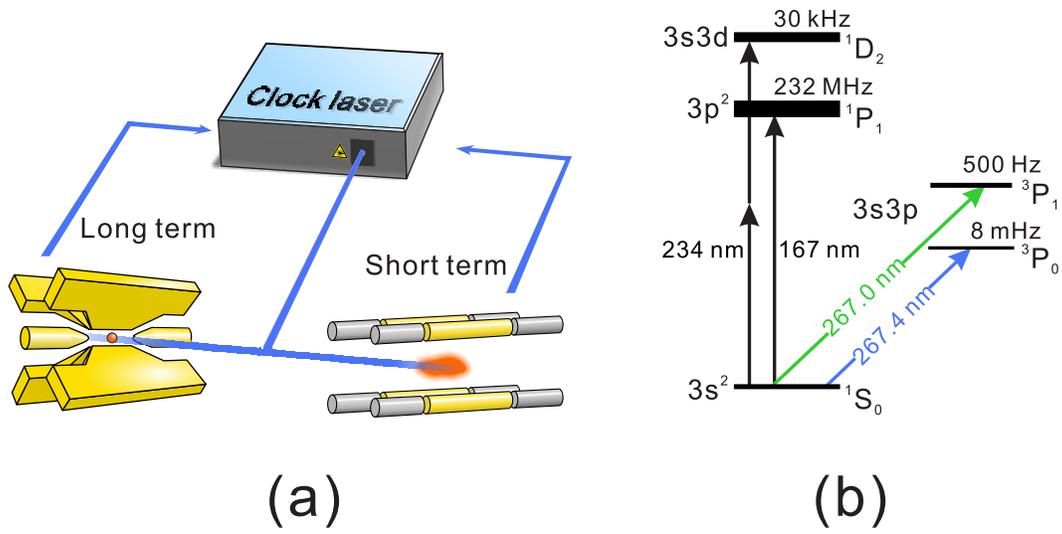}}
\caption{(Color online) (a) Proposed Al$^+$ ions optical clocks scheme. (b) Energy levels and relevant transitions of Al$^+$ ions.}
\label{fig1}
\end{figure}

\begin{figure}[tfb]
\centerline{\includegraphics[width=16cm]{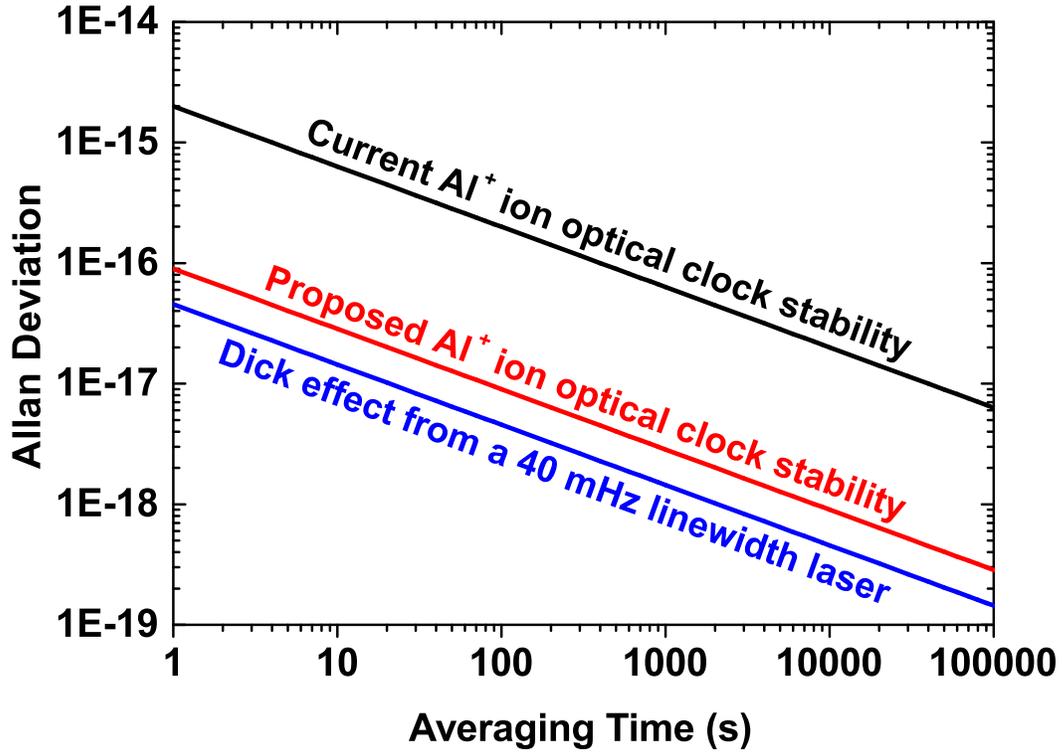}}
\caption{(Color online) Comparison of clock laser stability. The black line is the current Al$^+$ ion optical clock performance level of $2.0\times10^{-15}/\sqrt{\tau}$, and the red line is the expected clock laser stability improvement level of $9.0\times10^{-17}/\sqrt{\tau}$. For comparison, the Dick effect contribution from a 40 mHz line width laser (probe time 0.8 s, dead time 0.2 s) is shown as the blue line.}
\label{fig2}
\end{figure}

\end{document}